\documentstyle{article}
\newtheorem{property}{Property}
\newtheorem{lemma}{Lemma}

\newtheorem{definition}{Definition}
\newtheorem{corollary}{Corollary}
\begin{document}
\title{
\bf A small 1-way\\
quantum finite automaton
\thanks{This work was supported by Latvian Council of Science
under grant 96.0282.}}

\author{Arnolds \c{K}ikusts\\
{\small Department of Computer Science, University of Latvia},\\
{\small Rai\c{n}a bulv\={a}ris 29, LV-1459, Riga, Latvia}\\
{\small e-mail: sd70053@lanet.lv}}
\date{}
\maketitle

\begin{abstract}
We study 1-way quantum finite automata (QFAs)
and compare them with their classical counterparts.
We show that 1-way QFAs can be very space efficient.
We construct a 1-way QFAs that are quadratically smaller
than any equivalent deterministic finite automata
and give the correct answer with a large
probability by recognizing the languages $L_{n}$
in a two-letter alphabet $\{a,b\}$
for an arbitrary odd $n>2$:
$L_{n}=\{x\ |$\ in the word $x$ both the number
of letters $a$ and the number of letters $b$
are divisible by $n\}$.

\end{abstract}

\section{Introduction}
In [2] Kondacs and Watrous has shown that the class of languages recognized by 1-way
quantum finite automata (QFAs) is a proper subset of regular languages.
Nevertheless, in some situations 1-way QFAs are more powerful than equivalent
deterministic finite automata (DFAs).
In this paper, we continue the investigation of 1-way QFAs and
show the example where DFA needs considerably
more states than QFA.

Let $p$ be a prime.
Ambainis and Freivalds [1] consider the language
$L_{p}=\{a^{i}|i$ is divisible by $p\}$.
Any 1-way finite automaton recognizing $L_{p}$ has at least $p$ states.
In [1] is shown that there is an equivalent 1-way QFA
with $O(\log p)$ states.

In contrast to [1] we define an another language $L_{n}$
for all odd integers $n>2$:
$L_{n}=\{x\ |$\ in the word $x$ both the number
of letters $a$ and the number of letters $b$ are divisible by $n\}$.
It is easy to see that any 1-way finite automaton recognizing
$L_{n}$ has at least $n^{2}$ states.
We construct a 1-way QFA with $n+2$ states
accepting all words in $L_{n}$ with probability 1 and rejecting
words not in $L_{n}$ with probability less than $\frac{1}{k}$
where $k$ is the smaller prime multiplier in $n$.


\section{Quantum finite automata}
We consider 1-way quantum finite automata(QFA)
as defined in [2] (in [4] is given another definition of QFAs).

QFA is a tuple
$M=(Q;\Sigma ;\delta ;q_{0};Q_{acc};Q_{rej})$ where $Q$ is a finite set
of states, $\Sigma $ is an input alphabet, $\delta$ is a transition function,
$q_{0}\in Q$ is a starting state, and $Q_{acc}\subset Q$
and $Q_{rej}\subset Q$ are sets of accepting and rejecting states.
The states in $Q_{acc}$ and $Q_{rej}$ are called {\em halting states} and
the states in $Q_{non}=Q-(Q_{acc}\cup Q_{rej})$ are called
{\em non halting states}.
$\kappa$ and $\$$ are symbols that do not belong to $\Sigma$.
We use $\kappa$ and $\Sigma$ as the left and the right endmarker,
respectively. The {\em working alphabet} of
$M$ is $\Gamma = \Sigma \cup \{\kappa ;\$\}$.

A superposition of $M$ is any element of $l_{2}(Q)$ (the space of
mappings from $Q$ to $C$ with $l_{2}$ norm). For $q \in Q$,
$|q\rangle$ denotes the unit vector which takes value 1 at $q$
and 0 elsewhere. All elements of $l_{2}(Q)$ can be expressed as
linear combinations of vectors $|q\rangle$. We will use $\psi$ to
denote elements of $l_{2}(Q)$.

The transition function $\delta$ maps $Q\times \Gamma \times Q$
to $C$. The value $\delta (q_{1};a;q_{2})$ is the amplitude of
$|q_{2}\rangle$ in the superposition of states to which $M$ goes
from $|q_{1}\rangle$ after reading $a$. For $a\in \Gamma$, $V_{a}$
is a linear transformation on $l_{2}(Q)$ defined by
$$V_{a}(|q_{1}\rangle)=\sum\limits_{q_{2}\in Q}\delta(q_{1};a;q_{2}|q_{2}\rangle.$$
We require all $V_{a}$ to be unitary.

The computation of a QFA starts in the superposition $|q_{0}\rangle$.
Then transformations corresponding to the left endmarker $\kappa$,
the letters of the input word $x$ and the right endmarker $\$$ are
applied. The transformation corresponding to $a\in \Gamma$ consists
of two steps.

1. First, $V_{a}$ is applied. The new superposition $\psi^{\prime}$
is $V_{a}(\psi)$ where $\psi$ is the superposition before this step.

2. Then, $\psi^{\prime}$ is observed with respect to the observable
$E_{acc}\oplus E_{rej}\oplus E_{non}$ where
$E_{acc}=span\{|q_{1}\rangle :q\in Q_{acc}\}$,
$E_{rej}=span\{|q_{1}\rangle :q\in Q_{rej}\}$,
$E_{non}=span\{|q_{1}\rangle :q\in Q_{non}\}$.
This observation gives $x\in E_{i}$ with the probability equal to the
amplitude of the projection of $\psi^{\prime}$.
After that, the superposition collapses to this projection.\\

If we get $\psi^{\prime} \in E_{acc}$, the input is accepted.
If we get $\psi^{\prime} \in E_{rej}$, the input is rejected.
If we get $\psi^{\prime} \in E_{non}$, the next transformation is applied.\\

We regard these two transformations as reading a letter $a$.

\section{Preliminaries}

\begin{definition}
An arbitrary matrix in form
$$
\left (
\begin{array}{ccccc}
a_{0}&a_{1}&a_{2}&...&a_{n-1}\\
a_{n-1}&a_{0}&a_{1}&...&a_{n-2}\\
a_{n-2}&a_{n-1}&a_{0}&...&a_{n-3}\\
.&.&.&.&.\\
a_{1}&a_{2}&a_{3}&...&a_{0}
\end{array}
\right )
$$
is called a {\em shift matrix} and denoted
$(a_{0} \space  a_{1}   a_{2}  ...  a_{n-1})$.\\
\end{definition}

\begin{property}
Product of a shift matrix $A = (a_{0} \space  a_{1}   a_{2}  ...  a_{n-1})$
and a shift matrix $B = (b_{0} \space  b_{1}   b_{2}  ...  b_{n-1})$
is a shift matrix:\\
$  A\cdot B = (c_{0} \space  c_{1}   c_{2}  ...  c_{n-1}) $ where\\
$$ c_{i} = \sum\limits_{j=0}^{p-1} a_{j}\cdot b_{(i-j) \ \bmod \ n} $$
\end{property}

\begin{property}
Shift matrices commute.
\end{property}

\begin{property}
For all integers $c_{1}$ and $c_{2}$\\
$$ \sum\limits_{j=0}^{n-1} e^{\frac{2\pi}{n}{\rm i}\cdot c_{1}j^{2}} =
\sum\limits_{j=0}^{n-1} e^{\frac{2\pi}{n}{\rm i}\cdot c_{1}(j+c_{2})^{2}} $$
\end{property}

\begin{property}
If $t$ are not divisible by greatest common divisor of
$b$ and $n$ then
$$\sum\limits^{n-1}_{j=0} e^{\frac{2\pi}{n}{\rm i}(bj^{2}-2jt)}=0$$
\end{property}

Let / be an operation for numbers $a,b\in\{0,1,2,...,n-1\}$:
\begin{center}
$a/b (b\neq 0) = c$ if $ c\cdot b=a (\bmod n) $ and $c\in\{0,1,2,...,n-1\}$.
\end{center}


\begin{property}
We have\\
$1$. $ a/b=a\cdot (1/b) $. \\
$2$. $ 1/a+b=(1+a\cdot b)/(a\cdot b) $. \\
$3$. $ 1/(a/b)=b/a $.
\end{property}

\begin{definition}
A matrix $A$ is called unitary if $ A\cdot A^{\prime}=A^{\prime}\cdot A=I $
where $A^{\prime}$ is transpose complex conjugate of $A$ and $I$ is the
identity matrix.
\end{definition}


\begin{property}
Product of unitary matrices is a unitary matrix.
\end{property}

\begin{property}
In a unitary matrix the sum of squares of moduli of elements of each row
is equal to $1$.
\end{property}

\section{Main lemmas}
Let $M_{n}=(m_{0} m_{1} m_{2} ... m_{n-1})$ be a shift matrix where
$$m_{j}=\frac{1}{\sqrt{n}}e^{\frac{2\pi}{p}{\rm i}\cdot j^{2}},$$
$n=p_{1}p_{2}...p_{k}$ is an odd integer $>$ 2 where
$p_{1},p_{2},...,p_{k}$ is the prime multipliers of $n$.
Let $p_{min}$ be min$(p_{1},p_{2},...,p_{k})$.

\begin{property}
$M_{n}$ is a unitary matrix.
\end{property}

\begin{definition}
For an integer $l(n=l\cdot g)$
the unitary $n\times n$ shift matrix
$$(a_{0\cdot l} 0 ... 0 a_{1\cdot l} 0 ... 0 a_{2\cdot l} 0 ... 0
a_{(g-1)\cdot l} 0 ... 0)$$
where
$$a_{j\cdot l}=c\cdot e^{\frac{2\pi}{n}{\rm i}klj^{2}}
\hspace{40mm}
(1)$$\\
($c$ is a complex number and $k$ is an integer)
is called {\em special shift matrix}.
\end{definition}

\begin{property}
If $A$ is the special shift matrix then
$A^{\prime}$ is the special shift matrix
and $l$ of $A$ is equal to $l$ of $A^{\prime}$.
\end{property}

Let $A$ be a special shift matrix.

\begin{lemma}
$A\cdot M_{n}$ is a special shift matrix.
\end{lemma}

{\bf Proof.}
From $A\cdot M=(a_{0},a_{1} ... a_{n-1})$ where
$$a_{t}=\frac{c}{\sqrt{n}}\sum\limits^{g-1}_{j=0} e^{\frac{2\pi}{n}{\rm i}
(klj^{2}+(lj-t)^{2})}.$$
Let $d$ be the greatest common divisor of $k+l$ and $g$.\\
\\
{\em case 1.}\ $t$ is divisible by $d$.\\
Let $\frac{t}{d}$ be $t_{1}$ and $\frac{k+l}{d}$ be $k_{1}$.
We find $f$ that $lk_{1}f=l (\bmod n)$
(such $f$ exists because greatest common divisor of $k_{1}$ and $g$
is equal to 1).\\
Then
$a_{t}=\frac{c}{\sqrt{n}}\sum\limits^{g-1}_{j=0} e^{\frac{2\pi}{n}{\rm i}
d(k_{1}j^{2}-2ljt_{1}+dt_{1}^{2})}=
\frac{c}{\sqrt{n}}\sum\limits^{g-1}_{j=0} e^{\frac{2\pi}{n}{\rm i}
d(k_{1}j^{2}-2lk_{1}fjt_{1}+dt_{1}^{2})}=$\\
$\frac{c}{\sqrt{n}}\sum\limits^{g-1}_{j=0} e^{\frac{2\pi}{n}{\rm i}
d(k_{1}l(j-ft_{1})^{2}-lk_{1}f^{2}t_{1}^{2}+dt_{1}^{2})}=
\frac{c}{\sqrt{n}} e^{\frac{2\pi}{n}{\rm i} dt_{1}^{2}(d-lf)}
\sum\limits^{g-1}_{j=0} e^{\frac{2\pi}{n}{\rm i}(k+l)l(j-ft_{1})^{2}}$\\
From Property 3 we see that the expression
$\sum\limits^{g-1}_{j=0} e^{\frac{2\pi}{n}{\rm i}(k+l)l(j-ft_{1})^{2}}$
takes equal values for each $t$.
It means that each $a_{t}$ is in form (1).\\
\\
{\em case 2.}\ $t$ is not divisible by $d$.\\
$
a_{t}=\frac{c}{\sqrt{n}}\sum\limits^{g-1}_{j=0} e^{\frac{2\pi}{n}{\rm i}
(klj^{2}+l^{2}j^{2}-2ljt+t^{2})}=
a_{t}=\frac{c}{\sqrt{n}}e^{\frac{2\pi}{n}{\rm i}t^{2}}
\sum\limits^{g-1}_{j=0} e^{\frac{2\pi}{n}{\rm i}
(klj^{2}+l^{2}j^{2}-2ljt)}=
a_{t}=\frac{c}{\sqrt{n}}e^{\frac{2\pi}{n}{\rm i}t^{2}}
\sum\limits^{g-1}_{j=0} e^{\frac{2\pi}{g}{\rm i}
((k+l)j^{2}-2jt)}
$.\\
From Property 4 $a_{t}=0$.\\
\hspace*{118mm} $\Box$\\

It is easy to see that $M_{n}$ is the special shift matrix
where $l=1$, $g=n$, $k=1$ and $c=\frac{1}{\sqrt{n}}$.
From Lemma 1: each power of $M_{n}$ is a special shift matrix.

\begin{lemma}
If $n$ is a prime number then
$M^{n}_{n}$ is the special shift matrix with $l=n$
and for $0<s<p\ M^{s}_{n}$ is the special shift matrix with $l=1$.
\end{lemma}

{\bf Proof}.
Let $l_{s}$ be $l$ and $k_{s}$ be $k$ for the matrix $M^{s}(s<n)$.
By using Property 5 we will prove by induction over $j$ that
for $j<n$:\ $l_{j}=1$ and $$k_{j}=1/j.\ \ \ \ \ \  (2)$$
Induction basis: for $j=1$ holds $k_{1}=1/1=1/j$ and $l_{1}=1$.\\
Induction step:
$k_{j-1}=1/(j-1)(\bmod n)$ and $l_{j-1}=1$.
We must prove that $k_{j}=1/j$ and $l_{j}=1$.
We find the greatest common divisor $d$ of $k+1$ and $n$.
It is obvious that $d=1$($n$ is prime).
Hence $l_{j}=d=1$.
We find $f$ that $l_{j-1}(k_{j-1}+1)f=l_{j-1} (\bmod n)$.
It means that $f=1/(k_{j-1}+1)=1/(1/(j-1)+1)=(j-1)/j$
and $k_{j}=d-f=1-(j-1)/j=1/j$. \\
Now we must prove that $l_{n}=n$.
From (2) $k_{n-1}=1/(n-1)=n-1$, $k_{n-1}+1=n$.
$d$ is the greatest common divisor of $k+1$ and $n$.
It means that $l_{n}=d=n$.
\hspace*{118mm} $\Box$

\begin{lemma}
$M^{n}_{n}$ is the special shift matrix with $l=n$
and for $0<s<n\ M^{s}_{n}$ is the special shift matrix with $l<n$.
\end{lemma}

{\bf Proof}.
We prove this Lemma by induction over $h$ where
$h$ is the number of prime multipliers of $n$.

Induction basis: $h=1$.\\
It means that $n$ is prime, and the statement to be proved holds by Lemma 2.\\
Induction step: for all $n$ where $h=h_{1}$ holds the lemma.\\
Let us see what happens with $n$ where $h=h_{1}+1$.
Like in previous case is easy to prove that for $s<p_{min}$
$k_{s}=1/s$ and $l_{s}=1$.
We will prove that $k_{p_{min}}=1$ and $l_{p_{min}}=p_{min}$.
$k_{p_{min}-1}+1=p_{min}/(p_{min}-1)$. It means that $d=p_{min}=l_{p_{min}}$
and $f=p_{min}-1$ and $k_{p_{min}}=d-l_{p_{min}-1}f=d-f=1$.
It means that
$$M^{p_{min}}_{n}
=(a_{0\cdot l} 0 ... 0 a_{1\cdot l} 0 ... 0 a_{2\cdot l} 0 ... 0
a_{(g-1)\cdot l} 0 ... 0)\ \ \ (3)$$
where $l=p_{min}$ and $g=\frac{n}{p_{min}}$.
We know that for $h=h_{1}$ holds the lemma what
together with (3) implies $l_{n}=n$.
It is obvious that for $s<n$:\ $l_{n}\neq n$
(otherwise $M^{p_{min}\cdot c_{1}}_{n}\cdot M^{c_{2}}_{n}$
would be $c_{3}I$ where $c_{1},\ c_{2}$ are an integers and
$1\leq c_{2}< p_{min}$, $c_{3}$ is a complex number but from
Property 9 it would not be possible).
\hspace*{118mm} $\Box$

Let $M^{s}_{n}$ be $(x_{0}...x_{n-1})$.

Now from Lemma 3 and Property 7 follows:
\begin{corollary}
If $s=0(\bmod n) $ then $|x_{0}|^{2}=1$,
otherwise $|x_{0}|^{2}\leq \frac{1}{p_{min}}$.\\
\end{corollary}

Let $F$ be the $n\times n$ shift matrix $(0 1 0 0 ... 0)$.
\begin{property}
We have
$F^{2}=(0 0 1 0 ... 0)$,
$F^{3}=(0 0 0 1 0 ... 0)$,
...
$F^{n-1}=(0 0 0 0 ... 0 1)$,
and $F^{n}=(1 0 0 0 ... 0)=I$.
\end{property}

Let $C_{1}\cdot C_{2}\cdot C_{3}\cdot ... \cdot C_{a+b}$
be the shift matrix
$(x_{0} x_{1} x_{2} ... x_{n-1})$ where $C_{j}\in \{M_{n},F\}$, $a$ is
the number of matrices $M_{n}$ and $b$ is the number of matrices $F$.

Now from Corollary 1 and Property 2 and Property 10 follows
\begin{corollary}
\parbox[c]{14mm}{$|x_{0}|^{2}\hspace{2mm}\bigg\{\hspace{2mm}$}
\parbox[c]{100mm}{$=1$, if $a=0 (\bmod n)$ and $b=0 (\bmod n)$\\ $\leq \frac{1}{p_{min}}$, otherwise}\\
\end{corollary}

Let $m_{ij}$ be the elements of the matrix $M_{n}$
and $f_{ij}$ be the elements of the matrix $F$.

\section{Construction of the QFA}
We describe a 1-way QFA accepting the language $L_{n}$.
The automaton has $n+2$ states:
$q_{0}; q_{1}; ... q_{n-1}; q_{acc}$; and $q_{rej}$.
$Q_{acc}=\{q_{acc}\}$, $Q_{rej}=\{q_{rej}\}$.
The initial state is $|q_{0}\rangle$.
The transition function is
$$V_{a}(|q_{0}\rangle)=m_{0,0}|q_{0}\rangle+m_{0,1}|q_{1}\rangle+...+m_{0,n-1}|q_{n-1}\rangle;$$
$$V_{a}(|q_{1}\rangle)=m_{1,0}|q_{0}\rangle+m_{1,1}|q_{1}\rangle+...+m_{1,n-1}|q_{n-1}\rangle;$$
\ \ \ \ \ \ \ \ \ \ \ \ \ \ \ \ \ \ \ \ \ \ \ \ \ \ \ \ \ \ \ \ .
\ \ \ \ \ \ \ \ \ \ \ \ \ \ \ \ .
\ \ \ \ \ \ \ \ \ \ \ \ \ \ \ \ .
$$V_{a}(|q_{1}\rangle)=m_{n-1,0}|q_{0}\rangle+m_{n-1,1}|q_{1}\rangle+...+m_{n-1,n-1}|q_{n-1}\rangle;$$
$$V_{b}(|q_{0}\rangle)=f_{0,0}|q_{0}\rangle+f_{0,1}|q_{1}\rangle+...+f_{0,n-1}|q_{n-1}\rangle;$$
$$V_{b}(|q_{1}\rangle)=f_{1,0}|q_{0}\rangle+f_{1,1}|q_{1}\rangle+...+f_{1,n-1}|q_{n-1}\rangle;$$
\ \ \ \ \ \ \ \ \ \ \ \ \ \ \ \ \ \ \ \ \ \ \ \ \ \ \ \ \ \ \ \ .
\ \ \ \ \ \ \ \ \ \ \ \ \ \ \ \ .
\ \ \ \ \ \ \ \ \ \ \ \ \ \ \ \ .
$$V_{b}(|q_{1}\rangle)=f_{n-1,0}|q_{0}\rangle+f_{n-1,1}|q_{1}\rangle+...+f_{n-1,n-1}|q_{n-1}\rangle;$$
$$V_{\$}(|q_{0}\rangle)=|q_{acc}\rangle;
V_{\$}(|q_{1}\rangle)=|q_{rej}\rangle;
...
V_{\$}(|q_{n-1}\rangle)=|q_{rej}\rangle.$$
From Property 8 and Corollary 2 we see that the automaton works correctly.

\end{document}